# Physics and the Pythagorean Theorem

*By James Overduin[1,2] and Richard Conn Henry[2]*

[1]*Department of Physics, Astronomy and Geosciences, Towson University*
[2]*Department of Physics and Astronomy, Johns Hopkins University*

Pythagoras' theorem lies at the heart of physics as well as mathematics, yet its historical origins are obscure. We highlight a purely pictorial, gestalt-like proof that may have originated during the Zhou Dynasty. Generalizations of the Pythagorean theorem to three, four and more dimensions undergird fundamental laws including the energy-momentum relation of particle physics and the field equations of general relativity, and may hint at future unified theories. The intuitive, "pre-mathematical" nature of this theorem thus lends support to the Eddingtonian view that "the stuff of the world is mind-stuff".

The Pythagorean theorem $x^2 + y^2 = h^2$ (where *x,y* and *h* are the sides and hypotenuse of a right-angled triangle) is fundamental to physics as well as mathematics [Henry 2017]. Although widely credited to Pythagoras of Samos (c. 570-495 B.C.), versions of it were known many centuries earlier to Babylonian, Indian and Chinese mathematicians. New proofs have since been offered by so many people that it has been officially declared the world's "most proved theorem" in the *Guinness World Book of Records*. The list includes Euclid, Legendre, Leibniz, Huygens, Einstein, and a former U.S. President [Overduin, Molloy and Selway 2014].

Among the most interesting proofs are "gestalt-like" ones that can be grasped in a single glance using no mathematics at all. Such pictorial, tangram-like proofs may have originated in China [Anjing 1997]. One was described by Liu Hui in his 263 A.D. commentary on an earlier book, the *Jiuzhang suanshu* or Arithmetic in Nine Chapters [Wagner 1985]:

> The shorter leg multiplied by itself is the red square, and the longer leg multiplied by itself is the blue square. Let them be moved about so as to patch each other, each according to its type. Because the differences are completed, there is no instability. They form together the area of the square on the hypotenuse; extracting the square root gives the hypotenuse. [Qian Bacong 1963, 241]

Unfortunately, the figure that accompanied this proof has been lost to history. Donald Wagner has offered one possible reconstruction [Wagner 1985], but its intricacy is difficult to reconcile with the simplicity of the description above.

There is a simpler pictorial proof that accords better with Liu Hui's description (Fig. 1). The "shorter leg" here is $y$, and the "longer leg" is $x$. Multiplying each leg by itself gives the areas of the two "squares on the sides" (red and blue at left). Re-arranging the triangles, one sees immediately that this is the same as the area of the "square on the hypotenuse" $h$ (green at right).

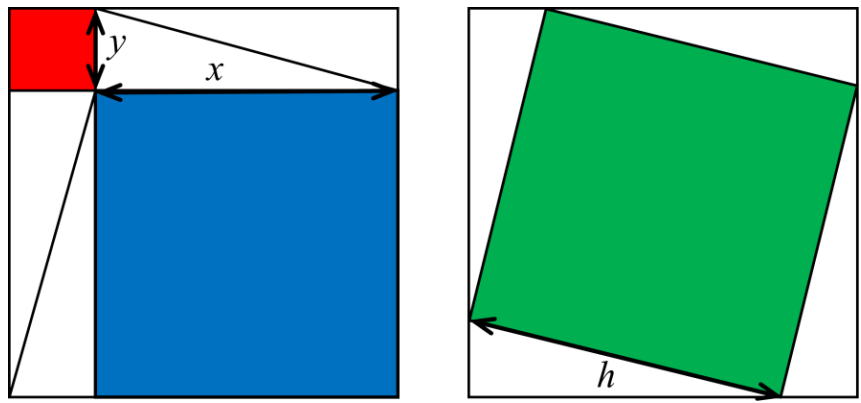

FIGURE 1: Gestalt-like proof of the Pythagorean theorem. By re-arranging the triangles (white), one sees immediately that the sum of the squares on the sides (red and blue) equals the square on the hypotenuse (green).

Pythagoras himself may also have discovered this proof [Maor 2007, Posamentier 2010]. However, its first published appearance seems to have been in a compilation by Elisha Scott Loomis [Loomis 1940]. Loomis in turn attributed it to Maurice Laisnez, "a high school boy, in the Junior-Senior High School of South Bend, Ind., and sent to me, May 16, 1939, by his class teacher, Wilson Thornton." It was later published in *Mathematics Magazine* [Isaac 1975]. It is reproduced in a book by Nelsen [Nelsen 1993], who attributes it to the *Zhoubi suanjing* [Arithmetical Classic of the Gnomon and the Circular Paths of Heaven], a classic of Chinese mathematics dating to the Zhou Dynasty (1046 – 256 B.C.). That document does contain a suggestive description similar to Liu Hui's [Needham and Wang 1959], but without the diagram. This proof is not only of historical interest, but is still used in mathematical physics today; see for instance Brill and Jacobson (2006).

Generalizing Pythagoras' theorem to three-dimensional space and applying it to small intervals of distance, we get $ds^2 = dx^2 + dy^2 + dz^2$. This allows us to describe distances along curved shapes using the power of calculus. If we make a giant leap of intuition and consider time as a kind of distance, we get Minkowski's generalization of the Pythagorean theorem,

$$ds^2 = dx^2 + dy^2 + dz^2 - c^2 dt^2 \quad , \tag{1}$$

where the constant $c$ is needed to convert units from seconds to meters. This is still just geometry, but it is also physics! Eq. (1) tells us that there is a special speed in the universe whose value must be independent of any observer's own motion along $x, y, z$. Impossible, but true! Moreover, thanks to Maxwell's discovery that $c = 1/\sqrt{\epsilon_0 \mu_0}$, we know that this quantity is not merely abstract, but the speed of an actual thing: three-dimensional vibrations in the electrical ($\epsilon_0$) and magnetic ($\mu_0$) fields, or "light."

But Eq. (1) does not just describe light; it is one of the foundations of the entire standard model of particle physics. For example, dividing through by $dt^2$ and defining "proper time" $\tau$ via $d\tau^2 \equiv -ds^2$, we get

$$(d\tau/dt)^2 = c^2 - \vec{v}^2 \quad , \tag{2}$$

where $\vec{v}^2 = v_x^2 + v_y^2 + v_z^2 = (dx/dt)^2 + (dy/dt)^2 + (dz/dt)^2$. Defining "energy" and "momentum" via $E \equiv \gamma m c^2$ and $\vec{p} \equiv \gamma m \vec{v}$ where $\gamma \equiv 1/\sqrt{1 - \vec{v}^2/c^2}$, we get

$$(mc^2)^2 = E^2 - (c\vec{p})^2 \quad . \tag{3}$$

Known as the energy-momentum relation, Eq. (3) is typically derived in relativity textbooks by combining energy and momentum in a single four-dimensional vector, the four-momentum $p^\alpha \equiv (E/c, \vec{p})$, and then showing that the scalar product of this vector with itself equals rest mass squared: $p^\alpha p_\alpha = (E/c)^2 - \vec{p}^2 = (mc)^2$ [see, e.g., Moore 2013]. Here $\alpha$ is an index labeling the component of any four-vector; its value ranges over 0 (for time) and 1,2,3 (for space). Any repeated index implies summation, so $p^\alpha p_\alpha = p^0 p_0 + p^1 p_1 + p^2 p_2 + p^3 p_3$. It is significant that the quantity on the right ($mc$) is *invariant:* it has the same value in all reference frames.

Equivalently, and more simply, we can see how Eq. (3) is just the Pythagorean theorem expressed in units of *energy* rather than distance (Fig. 2). The familiar expressions describing massive particles at rest ($E = mc^2$) and massless particles at the speed of light ($E = c|\vec{p}|$) are merely special cases of this more general physical law.

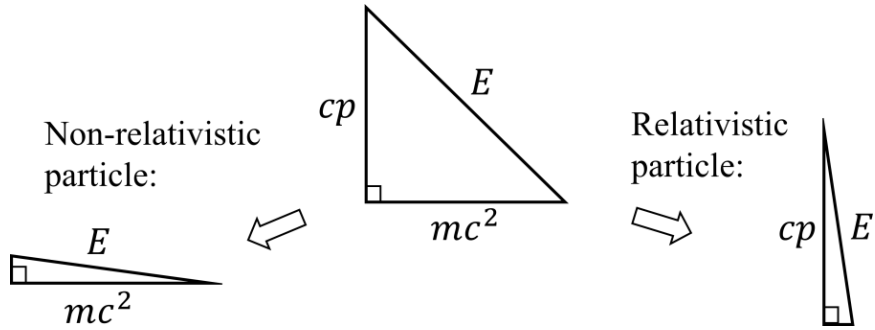

FIGURE 2: the Pythagorean theorem expressed in terms of energy

If we make another intuitive leap with Gauss and Riemann, we can allow movement in one direction to influence movement in the others and introduce coefficients in Eq. (1), so that

$$ds^2 = g_{xx}dx^2 + g_{xy}dxdy + g_{xz}dxdz - g_{xt}dxdt \\ + g_{yy}dy^2 + g_{yz}dydz - g_{yt}dydt \\ + g_{zz}dz^2 - g_{zt}dzdt \\ - g_{tt}dt^2 \quad . \quad (4)$$

Again this is just geometry, but it is physics too! The coefficients $g_{\alpha\beta}(x,y,z,t)$ describe the shape of the space and time in which we make our measurements. And thanks to Einstein, we know that this shape is not merely abstract. If any of the $g_{\alpha\beta}$ are different from one, we will feel it as *gravity*! The theory that follows from this insight, general relativity, is notoriously incompatible with the standard model of particle physics --- but we see here that both are ultimately based on Pythagoras' fertile theorem. (Strictly speaking, we should attach factors of two to the off-diagonal terms in Eq. (4), since nature presumably does not distinguish, for instance, between moving along $dx$ and then $dy$, or vice versa. Or if it does, we could generalize even further and allow all the $g_{\beta\alpha}$ to differ from $g_{\alpha\beta}$, an approach to the possible unification of forces that was originally investigated by Einstein with his assistant Ernst Straus and now goes by the name of nonsymmetric gravitational theory [Moffat 1995].)

It gets better. We need not stop with Minkowski. What if there are more fundamental degrees of freedom than just space and time? After all, there are three base quantities in physics: length, duration --- and *mass*. Imagine for a moment that what we measure as rest mass really labels distance along a fifth direction; then Eq. (1) generalizes to

$$dS^2 = dx^2 + dy^2 + dz^2 - c^2 dt^2 + \left(\frac{G}{c^2}\right)^2 dm^2 \qquad , \qquad (5)$$

where we use an upper-case "$S$" to denote five- instead of four-dimensional distance, and introduce Newton's gravitational constant ($G$) to convert kilograms to meters. Eq. (5) is bold! But no bolder than Minkowski's Eq. (1). It means that rest mass is not necessarily constant, as we may have believed (just as space and time are not the separate things we believed them to be before Minkowski). Particles can move in the mass direction in principle; this means their masses may vary. Before discarding the idea immediately, consider that any such variation will be almost impossible to detect, because of the values of $c$ and $G$. Relativistic effects depend on the Lorentz factor $\gamma$ defined above; with a masslike fifth dimension this becomes

$$\gamma_5 = \frac{1}{\sqrt{1 - \left(\frac{v}{c}\right)^2 - \left(\frac{G\,dm}{c^3 dt}\right)^2}} \qquad . \qquad (6)$$

To estimate the difference between $\gamma$ and $\gamma_5$ in a practical situation, consider for example an electron in the hydrogen ground state. It moves in the space direction at 1/130[th] of its speed in the time direction, $v/c = 0.0075$. What about the mass direction? The success of observational cosmology implies that its mass has probably not changed significantly since the time of cosmic nucleosynthesis, about ten minutes after the big bang. To get an upper limit, suppose that its entire rest mass was generated by some unknown process during those first ten minutes; then $Gdm/c^3 dt = 4 \times 10^{-69}$! If indeed something like Eq. (6) holds, then it probably describes a slow variation of particle rest masses throughout the universe over cosmological timescales [Bekenstein 1977, Liu and Wesson 2000, Wetterich 2014, Overduin and Ali 2017].

So Eq. (5) is plausible, if hard to test. Now consider how it beautifies physics. In standard (four-dimensional) relativity, Eq. (1) divides the universe into three separate regions. Matter lives and moves within the "light cone" defined by $ds^2 < 0$, or $|\vec{v}| < c$ according to Eq. (2). The region $ds^2 > 0$ is "elsewhen"; whatever exists there can neither act upon us nor be acted upon by us. Photons and other massless particles (gluons, gravitons) cling to the boundary between the two regions where $m = 0$ in Eq. (3), or $d\tau = 0$ according to Eq. (2).

That last statement is remarkable. To our three-dimensional senses, light appears to move very quickly, but in four-dimensional reality, it exists in a motionless state where no (proper) time ever passes. Indeed, insofar as light occupies no spacetime in four dimensions, one could debate whether it really "exists" at all.

*In five-dimensional relativity, that same statement may apply to everything!* To see this, repeat the derivation that led to Eq. (5) above, but with the new dimension included. In the limit where $dm/dt = 0$, you will recover the very same equation, but with the mass term simply moved to the right-hand side:

$$0 = E^2 - (c\vec{p})^2 - (mc^2)^2 \qquad . \qquad (7)$$

As before, $E$ arises from the time component of Pythagoras' theorem, and $\vec{p}$ with space. Now, however, $m$ is no longer the invariant magnitude of the momentum four-vector. In fact, it is not invariant at all, and does not divide the universe into causally distinct regions as before. Instead, it is merely the fifth *component* of a five-dimensional momentum vector $p^A \equiv (E/c, \vec{p}, mc)$, where the index labeled by $A$ ranges over 0 (for time), 1,2,3 (for space), *and 4* (for mass). The magnitude of this vector, $p^A p_A = (E/c)^2 - \vec{p}^2 - (mc)^2$, is even simpler invariant than mass: it is *zero!* Eq. (7) just states this explicitly. If it is correct, then from a five-dimensional point of view all matter lives in the same exalted state as light. In the language of relativity, *all particles follow null geodesics with $dS = 0$.*

The condition $dS = 0$ is not guaranteed in five-dimensional relativity, but is consistent with certain choices of coordinates, and seems to be logically favored in several ways [Seahra and Wesson 2001, Wesson 2008, Wesson 2009]. We might imagine the situation schematically with a generalization of Fig. 2 in which all three spatial directions are compressed into a single $\vec{x}$-axis (Fig. 3). Choosing coordinates such that $p^A p_A = 0$ (or equivalently, assuming $dS = 0$ for all particles) has the effect of "pushing" $E$ down into the $\vec{x}$-$t$ plane so that the five-dimensional Pythagorean energy theorem effectively agrees with the four-dimensional one and Eq. (3) is satisfied automatically, $E^2 = (mc^2)^2 + (c\vec{p})^2$.

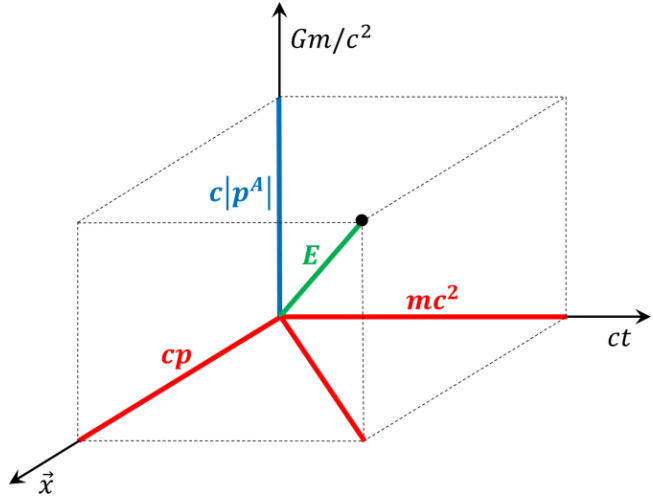

FIGURE 3: the Pythagorean theorem in five dimensions

Of course, even if we do not set $dS = 0$ in principle, the value of the invariant $p^A p_A$ may be undetectably small in practice. On dimensional grounds, we know that the "height" of the cube in Fig. 3 must be of order $c|p^A| \approx mc \, d/dt(Gm/c^2) = Gm\dot{m}/c$ where $|p^A|$ means the magnitude of $p^A$ and $\dot{m}$ is the time rate of change in mass. Adopting the same numbers as above, we find that $c|p^A| \lesssim 2 \times 10^{-63}$ eV for the electron.

There are other, deeper reasons to extend Pythagoras' theorem, and the theory of gravity based on it (general relativity), to higher dimensions. Einstein's gravitational field equations are obtained by differentiating the Pythagorean coefficients $g_{\alpha\beta}$ in Eq. (4) with respect to space and time and combining them in something called the Einstein tensor $G^{\alpha\beta}$. *Gravity is geometry!* These bumps and wrinkles in spacetime are then sourced by matter, as contained in the "energy-momentum tensor" $T^{\alpha\beta}$ on the right-hand side: $G^{\alpha\beta} = (8\pi G/c^2)T^{\alpha\beta}$. As John Wheeler put it, "matter tells spacetime how to curve, and curved spacetime tells matter how to move." Despite the undeniable beauty of this picture, it is well known that Einstein himself remained unhappy with its dualistic division of nature into "field" and "source." In 1936 he compared his field equations to a mansion, one wing of which is built of fine marble and the other of low-grade wood [Einstein 1936]; and he still echoed this point in a posthumous edition of *The Meaning of Relativity* twenty years later [Einstein 1956].

In five dimensions, this dichotomy disappears. The field equations read simply $G^{AB} = 0$. (The Einstein tensor is defined exactly as before, but derivatives now run over mass as well as space and time.) No matter: *everything* is geometry! When one extracts the four-dimensional $(\alpha, \beta)$ components of these equations, one recovers the full four-dimensional Einstein equations $G^{\alpha\beta} = (8\pi G/c^2)T^{\alpha\beta}$, with matter and energy ($T^{\alpha\beta}$) induced in four dimensions from *empty* five-dimensional spacetime. This is consistent with the idea expressed above, that matter is not fundamentally different from light; and with Einstein's dream of a theory unifying fields with their sources. He too experimented with higher dimensions, but was unwilling to grant them the degree of physical reality that could have led to new physics. For more on this subject, readers are directed to a short article [Wesson *et al.* 1996], a longer review [Overduin and Wesson 1997] or a recent comprehensive book [Wesson and Overduin 2018].

We have seen that all of physics (whether in two, three, four dimensions or possibly more) can be seen as applied geometry --- and this geometry is so axiomatic that it does not require mathematical proof, but can be apprehended at once, by direct human intuition. Could it be, as maintained by Arthur Eddington and others [Eddington 1958, Henry 2005, Wesson 2010], that "the stuff of the world is mind-stuff"? The historical roots and future implications of this wonderful idea deserve to be more fully explored.